# Moderate Deviations in Channel Coding

Yücel Altuğ, *Student Member, IEEE* and Aaron B. Wagner, *Member, IEEE*

**Abstract**

We consider block codes whose rate converges to the channel capacity with increasing block length at a certain speed and examine the best possible decay of the probability of error. We prove that a moderate deviation principle holds for all convergence rates between the large deviation and the central limit theorem regimes.

## I. INTRODUCTION

In block channel coding, there is a fundamental interplay between the *rate*, i.e., the amount of information transmitted per channel use, the *block length*, i.e., the total number of channel uses, and the *probability of error*. In this paper, we analyze the interplay between these three parameters for the best block codes. Specifically, we address the following question: for a given discrete memoryless channel, what is the fastest rate at which the error probability can decay to zero if the rate increases to the channel capacity with increasing block length? We begin by reviewing the literature on the interaction between these three basic parameters.

Shannon [1] formulated the channel coding problem and characterized the largest fixed rate such that the error probability could be driven to zero with increasing block length. Later, Strassen [2] considered the following more-refined characterization. Given a block length and an $\epsilon \in (0,1)$, what is the largest possible rate of a code with maximal error probability less than or equal to $\epsilon$? If $\epsilon \in (0, 1/2)$, then Strassen showed that this rate is equal to

$$\mathrm{C} - \frac{1}{\sqrt{n}} \sqrt{\sigma^2(W)} \Phi^{-1}(1-\epsilon) + \mathcal{O}\left(\frac{\log n}{n}\right),$$

where C denotes the channel capacity, $\Phi$ denotes the standard Gaussian distribution, and $\sigma^2(W)$ is a statistic of the channel defined later. More recently, Polyanskiy *et al.* [3], [4], provided an improved characterization of the $\mathcal{O}((\log n)/n)$ term and extended the result to Gaussian channels. Following the convention of [3], we call $\sigma^2(W)$ the *dispersion* of the channel. We note that although Strassen's result is classical, there is a renewed interest in his setup; see, e.g., [5]–[15], and references therein.

Another approach to the characterization of the interplay between rate, block length, and the probability of error is the so-called *error exponents*, which can be formulated as follows. Given a discrete memoryless channel and a fixed rate below the capacity[1], what is the best exponential rate of decay of the error probability with the block length? Classical results characterized the best exponent at rates close to capacity for a broad class of channels [16]–[22].

Our result lies between Strassen's result and error exponents in the sense that we require the rate to approach capacity and the error probability to simultaneously tend to zero. This formulation is arguably more relevant to practical code design than either error exponents or Strassen's result. The goal in channel coding is, after all, to attain a rate that is close to capacity *and* an error probability that is close to zero. Although error exponents allow for vanishing error probabilities, the rate is bounded away from capacity. In Strassen's result, on the other hand, the rate approaches capacity, but the error probability is bounded away from zero.

To place this formulation in context, it is helpful to consider the more-elementary setup of a sum of independent and identically distributed (i.i.d.) random variables. If we scale the sum with $1/n$, it converges to the mean by the law of large numbers. Cramér's Theorem (e.g. [23, Theorem 2.2.3]) characterizes the probability that the unnormalized sum makes an order-$n$ deviation from its mean. This probability decays exponentially in $n$, and Cramér's characterization of the exponent is now termed a *large deviations* result. The central limit theorem, on the other hand, characterizes the probability that the unnormalized sum makes an order-$\sqrt{n}$ deviation. As $n$ tends to infinity, this probability converges to a positive constant that is governed by the Normal distribution. Likewise, one can characterize the probability that the unnormalized sum makes a deviation whose size lies between these two extremes [23, Theorem 3.7.1] This is now called a *moderate deviations* result. Error exponents in channel coding are analogous to large deviations for i.i.d. sums, in that they both characterize exponentially small probabilities using similar techniques. Strassen's result is akin to the central limit theorem; indeed, it is sometimes called the *normal approximation*. The result in this paper is analogous to moderation deviations.

---

[1]In the literature, there is a considerable amount of work on the error exponents for rates above the capacity, not only for discrete memoryless channels, but also for various other problems, as well. However, we shall only be concerned with rates below the capacity here.

Although moderate deviations have been a fixture of probability theory for some time (e.g., [24]–[26], [27, Sec. XVI.7], [28, Chapter 8] and references therein), they appeared in the information theory literature only recently. The present result was first proven for positive discrete memoryless channels [29]. Prior to that, apparently the only moderate deviations result in information theory was the work of He *et al.* [30]–[33] on the Slepian-Wolf problem. Polyanskiy and Verdú [34] improved the result in [29] by relaxing the positivity assumption and extending it to Gaussian channels, among other contributions. More recently, moderate deviations in lossy source coding and hypothesis testing problems have been investigated by Tan [35] and Sason [36], respectively.

The result provided here improves upon the conference version [29] by relaxing the positivity assumption and simplifying the argument. The proof is different from that of Polyanskiy and Verdú, who rely on methods from [4] and powerful results from probability theory. It is also different from that of He *et al.* and Tan, who use type theory. It is worth noting that standard finite block length bounds on the rate and error probability are insufficient to obtain a conclusive moderate deviations result, and new bounds, such as those obtained with the aforementioned techniques, are necessary.

The organization of the paper is as follows. In Section II we define the relevant notions and state our result, Theorems 2.1 and 2.2. Section III-A cotains the proof of the direct part, and Section III-B contains the proof of the converse part.

**Notation:** Boldface letters denote vectors; regular letters with subscripts denote individual elements of vectors. Furthermore, capital letters represent random variables and lowercase letters denote individual realizations of the corresponding random variable. Throughout the paper, all logarithms are base-$e$. Given a finite set $\mathcal{X}$, $\mathcal{P}(\mathcal{X})$ denotes the set of all probability distributions defined on $\mathcal{X}$. Similarly, given two finite sets $\mathcal{X}$ and $\mathcal{Y}$, $\mathcal{P}(\mathcal{Y}|\mathcal{X})$ denotes the set of all stochastic matrices from $\mathcal{X}$ to $\mathcal{Y}$. Given any finite set $\mathcal{X}$ and for any $P \in \mathcal{P}(\mathcal{X})$, $\mathcal{S}(P)$ denotes the support of $P$. The sets $\mathbb{R}, \mathbb{R}_+$ and $\mathbb{R}^+$ denote real, non-negative real and positive real numbers, respectively. The set $\mathbb{Z}^+$ denotes positive integers. We follow the notation of Csiszár–Körner [21] for the fundamental information-theoretic notions.

## II. Definitions, Statement of the Main Result and the Auxiliary Results

### A. Definitions

Given $W \in \mathcal{P}(\mathcal{Y}|\mathcal{X})$, $(f, \varphi)$ denotes a *code*, with $f(\cdot)$ (resp. $\varphi(\cdot)$) being the encoding (resp. decoding) function. For a given code $(f, \varphi)$, $e_m(W, f, \varphi)$ denotes the *conditional probability of error* for message $m$, $e(W, f, \varphi)$ denotes the *maximal probability of error* and $\bar{e}(W, f, \varphi)$ denotes the *average probability of error*. $\mathrm{E}_\mathrm{o} : \mathbb{R}_+ \times \mathcal{P}(\mathcal{X}) \to \mathbb{R}$ denotes the function defined as

$$\mathrm{E}_\mathrm{o}(\rho, P) := -\log \sum_{y \in \mathcal{Y}} \left( \sum_{x \in \mathcal{X}} P(x) W(y|x)^{\frac{1}{1+\rho}} \right)^{1+\rho}, \tag{1}$$

for all $P \in \mathcal{P}(\mathcal{X})$ and $\rho \in \mathbb{R}_+$ (cf. [22, eq. (5.6.14)]). For any $R \in \mathbb{R}$, the random coding and sphere packing exponents, $\mathrm{E}_\mathrm{r}(R, W)$ and $\mathrm{E}_\mathrm{SP}(R, W)$, are defined as

$$\mathrm{E}_\mathrm{r}(R, W) := \max_{P \in \mathcal{P}(\mathcal{X})} \max_{0 \leq \rho \leq 1} \left\{ -\rho R + \mathrm{E}_\mathrm{o}(\rho, P) \right\}, \tag{2}$$

and

$$\mathrm{E}_\mathrm{SP}(R, W) = \max_{P \in \mathcal{P}(\mathcal{X})} \sup_{\rho \geq 0} \left\{ -\rho R + \mathrm{E}_\mathrm{o}(\rho, P) \right\}, \tag{3}$$

respectively. The following is a well-known result (e.g., [20, Theorem 18], [21, Ex. 2.5.23])

$$\mathrm{E}_\mathrm{SP}(R, W) = \max_{P \in \mathcal{P}(\mathcal{X})} \min_{V \in \mathcal{P}(\mathcal{Y}|\mathcal{X}) : \mathrm{I}(P;V) \leq R} \mathrm{D}(V \| W | P), \tag{4}$$

for any $R \in \mathbb{R}_+$.

Given any $W \in \mathcal{P}(\mathcal{Y}|\mathcal{X})$ and $P \in \mathcal{P}(\mathcal{X})$, we define

$$\sigma^2(P, W) := \mathrm{Var}_{P \times W}\left[\log \frac{W(Y|X)}{\sum_{z \in \mathcal{X}} P(z) W(Y|z)}\right]. \tag{5}$$

Using (5), we further define[2]

$$\sigma^2(W) := \min_{P \in \mathcal{P}(\mathcal{X}) : \mathrm{I}(P;W) = \mathrm{C}} \sigma^2(P, W), \tag{6}$$

and let $\tilde{P}(W)$ denote some element of $\mathcal{P}(\mathcal{X})$ that achieves the minimum in (6).

## B. Statement of the Main Result

The next two theorems comprise our main result.

*Theorem 2.1:* For any $W \in \mathcal{P}(\mathcal{Y}|\mathcal{X})$ with $\sigma^2(W) > 0$,[3] for any sequence of real numbers $\{\epsilon_n\}_{n \geq 1}$ satisfying

$$\begin{aligned}&\text{(i) } \epsilon_n \to 0, \text{ as } n \to \infty,\\ &\text{(ii) } \epsilon_n\sqrt{n} \to \infty, \text{ as } n \to \infty,\end{aligned} \tag{7}$$

there exists a sequence of codes $\{(f_n, \varphi_n)\}_{n \geq 1}$ that satisfies $R_n := \frac{\log |\varphi_n|}{n} \geq \mathrm{C} - \epsilon_n$, for all $n \in \mathbb{Z}^+$ and

$$\limsup_{n \to \infty} \frac{1}{n\epsilon_n^2} \log e(W, f_n, \varphi_n) \leq -\frac{1}{2\sigma^2(W)}. \tag{8}$$

*Theorem 2.2:* For any $W \in \mathcal{P}(\mathcal{Y}|\mathcal{X})$ with $\sigma^2(W) > 0$, for any sequence of real numbers $\{\epsilon_n\}_{n \geq 1}$ satisfying (7) and for any sequence of codes $\{(f_n, \varphi_n)\}_{n \geq 1}$ satisfying $R_n = \frac{\log |\varphi_n|}{n} \geq \mathrm{C} - \epsilon_n$, we have

$$\liminf_{n \to \infty} \frac{1}{n\epsilon_n^2} \log \bar{e}(W, f_n, \varphi_n) \geq -\frac{1}{2\sigma^2(W)}. \tag{9}$$

*Remark 2.1:* Polyanskiy and Verdú [34] show that the assumption $\sigma^2(W) > 0$ is necessary in order for

$$\frac{1}{n\epsilon_n^2} \log e(W, f_n, \varphi_n)$$

to have a finite limit. If $\sigma^2(W) = 0$, then we conjecture that

$$\frac{1}{n\epsilon_n} \log e(W, f_n, \varphi_n)$$

has a finite limit (see [34, Theorem 4]). $\diamond$

*Remark 2.2:* Our achievability proof follows from Gallager's random coding bound (e.g. [22, Corollary 2, pg. 140]), which states that for any rate $R$ and block length $n$, there exists an $(n, R)$ code $(f, \varphi)$ such that

$$e(W, f, \varphi) \leq 4 e^{-n\mathrm{E}_\mathrm{r}(R, W)}. \tag{10}$$

Since $n$ and $R$ are arbitrary, we can let $R = \mathrm{C} - \epsilon_n$ and approximate $\mathrm{E}_\mathrm{r}(\cdot, W)$ around C via a Taylor series to obtain Theorem 2.1. This line of reasoning is made rigorous in Section III-A.

The achievability argument is deceptively simple in that it obscures issues that must be confronted when proving the converse. To prove the converse, we would like to show that for any $\epsilon_n$ satisfying the hypothesis of the theorem and any $\alpha > 1$, there exists sequences $\beta_n$ and $\gamma_n$ satisfying

$$\frac{\beta_n}{\epsilon_n} \to 0 \tag{11}$$

$$\frac{1}{n\epsilon_n^2} \log \gamma_n \to 0, \tag{12}$$

---

[2] The minimum is well-defined owing to the continuity of $\sigma^2(\cdot, W)$ (cf. Remark 2.3).
[3] Since $\sigma^2(W) > 0$ implies that $\mathrm{C} > R_\infty(W) \geq 0$ (e.g. [22, pg. 160]) we have $\mathrm{C} > 0$.

such that for all sufficiently large $n$ and all $(n, \mathrm{C} - \epsilon_n)$ codes $(f, \varphi)$, we have

$$e(W, f, \varphi) \geq \gamma_n e^{-n\alpha \mathrm{E}_{\mathrm{SP}}(\mathrm{C} - \epsilon_n - \beta_n, W)}. \tag{13}$$

If one could prove such a bound, then one could obtain Theorem 2.2 by expanding $\mathrm{E}_{\mathrm{SP}}(R, W)$ as a Taylor series around $R = \mathrm{C}$ and taking the appropriate limit.

But it is not clear whether a bound like (13) holds. The authors' recent refinement of the classical sphere-packing bound [37, Theorem 2.1] establishes that for all $\epsilon > 0$, all fixed rates $R$ below capacity, and all sufficiently large $N$, any constant composition[4] $(n, R)$ code $(f, \varphi)$ satisfies

$$e(W, f, \varphi) \geq \frac{K(R)}{\sqrt{n}} \exp\left\{-n\mathrm{E}_{\mathrm{SP}}\left(R - \frac{(1+\epsilon)\log\sqrt{n}}{n}, W\right)\right\}. \tag{14}$$

Moreover, the $n$-dependence on the right side is essentially the best possible for a fixed $R$ [39].

Although the rate backoff in this bound clearly satisfies (11), whether the pre-factor satisfies (12) hinges $R$ dependence of $K(R)$. This dependence is not currently known, but it can be postulated via the following reasoning. In Strassen's regime, in which the rate approaches capacity at a speed of $1/\sqrt{n}$, the error probability is asymptotically constant [2], and a Taylor series expansion of the sphere-packing exponent shows that the exponential factor in (14) is also asymptotically constant in this regime. If we assume that (14) holds in this regime, then it follows that the pre-factor must also be asymptotically constant, which suggests that $K(R)$ might behave as $1/(\mathrm{C} - R)$. If this is true, then the pre-factor would satisfy (12), so (13) would hold.

We show that (13) indeed holds, although our proof does not involve characterizing how $K(R)$ varies with $R$.[5] Instead we prove (13) directly using a particular set of classical information theory results, which do not appear to have been used in combination before, to prove a version of the sphere-packing exponent that is especially tight at finite block lengths and rates near capacity. The fact that our proof is similar to existing derivations of the sphere-packing exponent and uses well-known ingredients might give the impression that the result is routine. In fact, the required bounds are quite delicate, as the above discussion illustrates, and many conceptually-similar approaches to proving the sphere-packing exponent fail to give a conclusive moderate deviations result. $\diamond$

### C. Auxiliary Results

*Lemma 2.1:* Given any $W \in \mathcal{P}(\mathcal{Y}|\mathcal{X})$ with no all-zero column, $\mathrm{E}_{\mathrm{o}}(\rho, P)$ possesses the following properties:
1) Given any $P \in \mathcal{P}(\mathcal{X})$, $\mathrm{E}_{\mathrm{o}}(\rho, P)$ is concave in $\rho \in \mathbb{R}_+$.
2) Given any $P \in \mathcal{P}(\mathcal{X})$,
$$\left.\frac{\partial \mathrm{E}_{\mathrm{o}}(\rho, P)}{\partial \rho}\right|_{\rho=0} = \mathrm{I}(P; W). \tag{15}$$

3) Given any $P \in \mathcal{P}(\mathcal{X})$,
$$\left.\frac{\partial^2 \mathrm{E}_{\mathrm{o}}(\rho, P)}{\partial \rho^2}\right|_{\rho=0} = -\sigma^2(P, W). \tag{16}$$

4) Given any $P \in \mathcal{P}(\mathcal{X})$,
$$\frac{\partial \mathrm{E}_{\mathrm{o}}(\rho, P)}{\partial \rho} \leq \mathrm{I}(P; W), \forall \rho \in \mathbb{R}_+. \tag{17}$$

5) $\frac{\partial \mathrm{E}_{\mathrm{o}}(\rho, P)}{\partial \rho}$ is continuous over $(\rho, P) \in \mathbb{R}_+ \times \mathcal{P}(\mathcal{X})$.
6) $\frac{\partial^2 \mathrm{E}_{\mathrm{o}}(\rho, P)}{\partial \rho^2}$ is continuous over $(\rho, P) \in \mathbb{R}_+ \times \mathcal{P}(\mathcal{X})$.
7) $\frac{\partial^3 \mathrm{E}_{\mathrm{o}}(\rho, P)}{\partial \rho^3}$ is continuous over $(\rho, P) \in \mathbb{R}_+ \times \mathcal{P}(\mathcal{X})$.

*Proof:* The proof is given in the Appendix A. ∎

*Remark 2.3:* Note that for any given $W \in \mathcal{P}(\mathcal{Y}|\mathcal{X})$, $\sigma^2(\cdot, W)$ is continuous on $\mathcal{P}(\mathcal{X})$, owing to items 3) and 6) of Lemma 2.1. $\diamond$

---

[4] If the channel is symmetric, then the constant composition assumption can be dropped (cf. [38]).
[5] Determining how $K(R)$ varies with $R$ is an interesting subject for future work.

## III. PROOF OF THE RESULTS

### A. *Proof of Theorem 2.1*

Let $W \in \mathcal{P}(\mathcal{Y}|\mathcal{X})$ be an arbitrary stochastic matrix satisfying the conditions stated in the theorem. Without loss of generality, suppose that $W$ has no all-zero columns. Further, let $\{\epsilon_n\}_{n \geq 1}$ be an arbitrary sequence of real numbers, satisfying (7). By (7) and the fact that $C > 0$, we have

$$C - \epsilon_n > 0, \tag{18}$$

for all sufficiently large $n$. Next, fix such an $n$. Gallager's random coding bound (e.g. [22, Corollary 2, pg. 140]) implies that there exists $(f_n, \varphi_n)$, such that $R_n \geq R_n := C - \epsilon_n$ and

$$e(W, f_n, \varphi_n) \leq 4 \exp\left\{-n \left[\max_{0 \leq \rho \leq 1} \{E_o(\rho, P) - \rho R_n\}\right]\right\}, \tag{19}$$

for all $P \in \mathcal{P}(\mathcal{X})$. Therefore (19) implies the existence of a sequence of codes $\{(f_n, \varphi_n)\}_{n \geq 1}$, s.t. for all $n \in \mathbb{Z}^+$, $R_n \geq C - \epsilon_n$ and

$$\frac{1}{n\epsilon_n^2} \log e(W, f_n, \varphi_n) \leq \frac{\log 4}{n\epsilon_n^2} - \frac{1}{\epsilon_n^2} \max_{0 \leq \rho \leq 1} \{E_o(\rho, P) - \rho R_n\}, \tag{20}$$

for all sufficiently large $n$ and any $P \in \mathcal{P}(\mathcal{X})$. Hence, it suffices to prove that (8) holds for this particular sequence of codes in order to conclude the result.

Using Taylor's Theorem, along with (15) and (16) (cf. items 2) and 3) of Lemma 2.1), for any $\rho \in \mathbb{R}_+$, we have

$$E_o(\rho, \tilde{P}(W)) = \rho C - \frac{\rho^2}{2} \sigma^2(W) + \frac{\rho^3}{6} \left.\frac{\partial^3 E_o(\rho, \tilde{P}(W))}{\partial \rho^3}\right|_{\rho = \bar{\rho}}, \tag{21}$$

for some $\bar{\rho} \in [0, \rho]$. Next, let $\rho_n = \frac{\epsilon_n}{\sigma^2(W)}$, for all $n \in \mathbb{Z}^+$. Then (21) yields,

$$\max_{0 \leq \rho \leq 1} \left\{E_o(\rho, \tilde{P}(W)) - \rho R_n\right\} \geq \frac{\epsilon_n^2}{2\sigma^2(W)} - \frac{\epsilon_n^3}{6\sigma^6(W)} \left| \left.\frac{\partial^3 E_o(\rho, \tilde{P}(W))}{\partial \rho^3}\right|_{\rho = \bar{\rho}_n} \right|, \tag{22}$$

for all sufficiently large $n$ and for some $\bar{\rho}_n \in [0, \rho_n]$.

Next, note that $\rho_n \leq 1$, for all sufficiently large $n$, since $\lim_{n \to \infty} \epsilon_n = 0$ (cf. (i) of (7)) and $\sigma^2(W) > 0$. We define

$$M := \max_{(\rho, P) \in [0,1] \times \mathcal{P}(\mathcal{X})} \left|\frac{\partial^3 E_o(\rho, P)}{\partial \rho^3}\right|. \tag{23}$$

Owing to item 7) of Lemma 2.1, the maximum in (23) is well-defined and finite. Therefore, (22) and (23) imply that

$$\max_{0 \leq \rho \leq 1} \left\{E_o(\rho, \tilde{P}(W)) - \rho R_n\right\} \geq \frac{\epsilon_n^2}{2\sigma^2(W)} - \frac{\epsilon_n^3}{6\sigma^6(W)} M, \tag{24}$$

for all sufficiently large $n$.

Substituting (24) into (20) yields

$$\frac{1}{n\epsilon_n^2} \log e(W, f_n, \varphi_n) \leq \frac{\log 4}{n\epsilon_n^2} - \frac{1}{2\sigma^2(W)} \left(1 - M \frac{\epsilon_n}{3\sigma^4(W)}\right), \tag{25}$$

which, in turn, implies (recall (7) and (23))

$$\limsup_{n \to \infty} \frac{1}{n\epsilon_n^2} \log e(W, f_n, \varphi_n) \leq -\frac{1}{2\sigma^2(W)},$$

which is (8) and hence we conclude the proof.

## B. Proof of Theorem 2.2

Let $W$ and $\{\epsilon_n\}_{n\geq 1}$ be as in Section III-A. Further, let $\{(f_n, \varphi_n)\}_{n\geq 1}$ be an arbitrary sequence of codes with $\frac{\log |\varphi_n|}{n} := R_n \geq C - \epsilon_n$, for all $n \in \mathbb{Z}^+$. Observe that owing to standard arguments used to switch from the maximum to average error probability (e.g. [18, eq. (4.41)]), it is sufficient to show the conclusion for the maximum error probability, i.e.,

$$\liminf_{n\to\infty} \frac{1}{n\epsilon_n^2} \log e(W, f_n, \varphi_n) \geq -\frac{1}{2\sigma^2(W)}, \tag{26}$$

in order to prove (9). By similar reasoning [21, pg. 171], we can assume that the code is constant composition.

Next, we briefly outline the rest of the proof, which consists of three steps. The first step is to prove a strong converse theorem, Lemma 3.1, tailored to the particular situation at hand. The second step is to use Lemma 3.1 and "change of measure" to prove (13) (cf. Remark 2.2). The final step is to approximate the exponent in (13) via a Taylor series to conclude the result.

*Remark 3.1:* Lemma 3.1, which could be of independent interest, is derived from Wolfowitz's converse to the channel coding theorem [40]. Although our version requires that the code be constant composition, an assumption not required by Wolfowitz, it shows that the error probability must be near unity if the rate exceeds the mutual information induced by the code. Wolfowitz requires the rate to exceed capacity. ◇

*Remark 3.2:* One of the well-known change of measure arguments is Marton's [41, eq. (12)]. Although Marton originally applied it to rate distortion, the application to channel coding is obvious. It does not seem sufficient to prove (13), however. Instead, we use a change of measure argument based on the log-sum inequality, given by Csiszár and Körner [21, pg. 167]. ◇

Define the constant $A$ as follows:

$$A := \max_{(P \times V) \in \mathcal{P}(\mathcal{X}) \times \mathcal{P}(\mathcal{Y}|\mathcal{X})} \mathrm{Var}\left[\log \frac{V(Y|X)}{Q(Y)}\right] + 1, \tag{27}$$

where $Q(y) := \sum_{x \in \mathcal{X}} P(x) V(y|x)$, $\forall y \in \mathcal{Y}$. Note that, since the cost function is continuous in the optimization variable and we work with finite alphabets, the maximum in (27) is well-defined and finite.

*Lemma 3.1:* (Strong Converse). Let $(f, \varphi)$ be an arbitrary constant composition code with block length n, common type $P$, and rate $R > 0$. Let $V \in \mathcal{P}(\mathcal{Y}|\mathcal{X})$ be an arbitrary stochastic matrix satisfying $I(P;V) \leq R - 2\delta$, for some $\delta > 0$. Then, we have

$$\bar{e}(V, f, \varphi) \geq 1 - \frac{A}{n\delta^2} - e^{-n\delta}, \tag{28}$$

where $A$ is defined in (27).

*Proof:* The proof is given in Appendix B. ∎

Next, fix some $0 < \gamma < 1/2$. Let $\psi \in \mathbb{R}^+$ be defined as

$$\psi^2 := \frac{2A}{\gamma}. \tag{29}$$

Note that for all sufficiently large $n$,

$$0 < C - \left(\epsilon_n + \frac{2\psi}{\sqrt{n}}\right), \tag{30}$$

$$e^{-\psi\sqrt{n}} \leq \gamma/2. \tag{31}$$

As a direct consequence of the Strong Converse lemma (with the choice of $\delta = \psi/\sqrt{n}$), for any $V \in \mathcal{P}(\mathcal{Y}|\mathcal{X})$ satisfying $I(P_n; V) \leq R_n - \frac{2\psi}{\sqrt{n}}$, we have

$$\exists m \in \mathcal{M} = \left\{1, \ldots, \lceil 2^{nR_n} \rceil\right\}, \text{ s.t. } e_m(V, f_n, \varphi_n) \geq 1 - \gamma, \tag{32}$$

for all sufficiently large $n$, such that (30) and (31) hold. Note that $n$ does not depend on the specific choice of $V$. Fix a sufficiently large $n$ such that (30) and (31) hold.

*Lemma 3.2:* (Change of Measure). Let $(f, \varphi)$ be an arbitrary constant composition code with block length $n$ and common type $P_n$. Then

$$e(W, f_n, \varphi_n) \geq \exp\left\{-n\left(\min_{V \in \mathcal{P}(\mathcal{Y}|\mathcal{X}) : I(P_n;V) \leq R_n - \frac{2\psi}{\sqrt{n}}} \left\{\frac{D(V||W|P_n)}{1-\gamma} + \frac{h(1-\gamma)}{n(1-\gamma)}\right\}\right)\right\}, \tag{33}$$

for all sufficiently large $n$ such that (30) and (31) hold, where $h(\cdot)$ is the binary entropy function, i.e. $h(p) := p\log(1/p) + (1-p)\log(1/(1-p))$, $\forall p \in [0,1]$.

*Proof:* The argument is due to Csiszár and Körner (e.g. [21, pg. 167]), and we state it for the sake of completeness. Fix $n$ and let $V$ be any channel such that

$$I(P_n; V) \leq R_n - \frac{2\psi}{\sqrt{n}}.$$

By the log-sum inequality (e.g. [21, pg. 48]), for any message $m$, we have

$$V^n(\varphi^{-1}(m)|\mathbf{x}^n(m))\log\frac{V^n(\varphi^{-1}(m)|\mathbf{x}^n(m))}{W^n(\varphi^{-1}(m)|\mathbf{x}^n(m))} + V^n((\varphi^{-1}(m))^c|\mathbf{x}^n(m))\log\frac{V^n((\varphi^{-1}(m))^c|\mathbf{x}^n(m))}{W^n((\varphi^{-1}(m))^c|\mathbf{x}^n(m))}$$
$$\leq D(V^n\|W^n|\mathbf{x}^n(m)),$$

where $\varphi^{-1}(m)$ denotes the decoding region for the $m$-th message and $(\varphi^{-1}(m))^c$ denotes its complement. This, in turn, implies that

$$V^n((\varphi^{-1}(m))^c|\mathbf{x}^n(m))\log\frac{1}{W^n(\varphi^{-1}(m)|\mathbf{x}^n(m))} \leq D(V^n\|W^n|\mathbf{x}^n(m)) + h(V^n(\varphi^{-1}(m)|\mathbf{x}^n(m))). \quad (34)$$

Applying this inequality to a message satisfying (32) gives (33). ∎

Equation (33), along with (4), implies that

$$e(W, f_n, \varphi_n) \geq e^{-\frac{h(1-\gamma)}{(1-\gamma)}} \exp\left\{-n\left(\frac{E_{SP}(C-\delta_n, W)}{1-\gamma}\right)\right\}, \quad (35)$$

where

$$\delta_n := \epsilon_n\left(1 + \frac{2\psi}{\sqrt{n\epsilon_n^2}}\right),$$

for all $n \in \mathbb{Z}^+$. Note that this establishes (13). We define

$$\alpha_n := 1 + \frac{2\psi}{\sqrt{n\epsilon_n^2}}, \ \forall n \in \mathbb{Z}^+,$$

and note that since $\epsilon_n\sqrt{n} \to \infty$ as $n \to \infty$ (cf. item (ii) of (7)), $\alpha_n \to 1$ as $n \to \infty$. Therefore, $\delta_n \to 0$ as $n \to \infty$ (cf. item (i) of (7)).

The third and final step of the proof is to approximate the exponent on the right side of (35). To this end, first note that if the rate is above the critical rate[6], i.e. $R \geq R_{cr}$, then $E_{SP}(R,W) = E_r(R,W)$ (e.g. [22, pg. 160]), which, in turn, implies that

$$E_{SP}(R,W) = E_r(R,W) = \max_{P \in \mathcal{P}(\mathcal{X})} \max_{0 \leq \rho \leq 1} \{-\rho R + E_o(\rho, P)\}, \quad (36)$$

by recalling (2).

Further, since $\sigma^2(W) > 0$, one can infer that (e.g. [22, pg. 160]) $R_{cr} < C$ and hence for all sufficiently large $n$, $C - \delta_n > R_{cr}$. This observation, coupled with (36), ensures that for all sufficiently large $n$, we have

$$E_{SP}(C-\delta_n, W) = E_r(C-\delta_n, W) = \max_{P \in \mathcal{P}(\mathcal{X})} \max_{0 \leq \rho \leq 1} \{-\rho[C - \delta_n] + E_o(\rho, P)\}. \quad (37)$$

*Proposition 3.1:* (Sphere–packing exponent around C)

$$\limsup_{n \to \infty} \frac{E_{SP}(C-\delta_n, W)}{\delta_n^2} \leq \frac{1}{2\sigma^2(W)}. \quad (38)$$

*Proof:* The proof is given in the Appendix C. ∎

---

[6] See [22, pg. 160] for the definition of $R_{cr}$.

Equipped with Proposition 3.1, we conclude the proof as follows. Recall that $\delta_n = \epsilon_n \alpha_n$, where $\alpha_n > 0$, for all $n \in \mathbb{Z}^+$ and $\alpha_n \to 1$ as $n \to \infty$. Hence,

$$\limsup_{n \to \infty} \frac{E_{\text{SP}}(C - \delta_n, W)}{\delta_n^2} = \limsup_{n \to \infty} \frac{E_{\text{SP}}(C - \delta_n, W)}{\epsilon_n^2}. \tag{39}$$

Since $\lim_{n \to \infty} n\epsilon_n^2 = \infty$ (cf. item (ii) of (7)), (35), (37) and (39) imply that

$$\liminf_{n \to \infty} \frac{1}{n\epsilon_n^2} \log e(W, f_n, \varphi_n) \geq -\frac{1}{2\sigma^2(W)} \frac{1}{1 - \gamma}. \tag{40}$$

Since $0 < \gamma < 1/2$ is arbitrary, letting $\gamma \to 0$ in the right side of (40) yields (26), which was to be shown.

## APPENDIX A
### PROOF OF LEMMA 2.1

Consider any $W \in \mathcal{P}(\mathcal{Y}|\mathcal{X})$. For all $y \in \mathcal{Y}$, define

$$\mathcal{X}_y := \{x \in \mathcal{X} : W(y|x) > 0\}. \tag{41}$$

Observe that owing to the no all-zero column assumption on $W$ and (41), for all $y \in \mathcal{Y}$, $\mathcal{X}_y \neq \emptyset$. Moreover, for any $P \in \mathcal{P}(\mathcal{X})$, there exists $y \in \mathcal{Y}$ with $\mathcal{X}_y \cap \mathcal{S}(P) \neq \emptyset$.

For all $y \in \mathcal{Y}$, define

$$f_y : \mathbb{R}_+ \times \mathcal{P}(\mathcal{X}) \to \mathbb{R}_+, \text{ s.t. } f_y(\rho, P) := \sum_{x \in \mathcal{X}} P(x) W(y|x)^{\frac{1}{(1+\rho)}}, \forall (\rho, P) \in \mathbb{R}_+ \times \mathcal{P}(\mathcal{X}). \tag{42}$$

Evidently $f_y(\cdot, \cdot)$ is continuous on $\mathbb{R}_+ \times \mathcal{P}(\mathcal{X})$. Also, straightforward calculation reveals that

$$\frac{\partial f_y(\rho, P)}{\partial \rho} = -\frac{1}{(1+\rho)^2} \sum_{x \in \mathcal{X}_y} P(x) W(y|x)^{\frac{1}{(1+\rho)}} \log W(y|x), \tag{43}$$

$$\frac{\partial^2 f_y(\rho, P)}{\partial \rho^2} = \frac{1}{(1+\rho)^3} \sum_{x \in \mathcal{X}_y} P(x) W(y|x)^{\frac{1}{(1+\rho)}} \log W(y|x) \left[2 + \frac{\log W(y|x)}{(1+\rho)}\right], \tag{44}$$

$$\frac{\partial^3 f_y(\rho, P)}{\partial \rho^3} = -\frac{1}{(1+\rho)^4} \sum_{x \in \mathcal{X}_y} P(x) W(y|x)^{\frac{1}{(1+\rho)}} \log W(y|x) \left[6 + \frac{6 \log W(y|x)}{(1+\rho)} + \frac{(\log W(y|x))^2}{(1+\rho)^2}\right]. \tag{45}$$

Further,

$$\forall P \in \mathcal{P}(\mathcal{X}), \text{ s.t. } \mathcal{S}(P) \cap \mathcal{X}_y = \emptyset, \ f_y(\cdot, P) = 0. \tag{46}$$

Equation (46), coupled with (43), (44) and (45), implies that $\frac{\partial f_y(\rho, P)}{\partial \rho}$, $\frac{\partial^2 f_y(\rho, P)}{\partial \rho^2}$ and $\frac{\partial^3 f_y(\rho, P)}{\partial \rho^3}$ are continuous for all $(\rho, P) \in \mathbb{R}_+ \times \mathcal{P}(\mathcal{X})$.

For all $y \in \mathcal{Y}$, define

$$g_y : \mathbb{R}_+ \times \mathcal{P}(\mathcal{X}) \to \mathbb{R}_+, \text{ s.t. } g_y(\rho, P) := f_y(\rho, P)^{(1+\rho)}, \tag{47}$$

where $f_y(\cdot, \cdot)$ is defined in (42). It follows that $g_y(\cdot, \cdot)$ is continuous on $\mathbb{R}_+ \times \mathcal{P}(\mathcal{X})$.

Note that

$$\forall P \in \mathcal{P}(\mathcal{X}), \text{ s.t. } \mathcal{S}(P) \cap \mathcal{X}_y = \emptyset, \ g_y(\cdot, P) = 0. \tag{48}$$

Consider any $P \in \mathcal{P}(\mathcal{X})$ with $\mathcal{S}(P) \cap \mathcal{X}_y \neq \emptyset$. By noting $g_y(\rho, P) = e^{(1+\rho) \log f_y(\rho, P)}$, one can check that

$$\frac{\partial g_y(\rho, P)}{\partial \rho} = g_y(\rho, P) \left[(1+\rho) \frac{\frac{\partial f_y(\rho, P)}{\partial \rho}}{f_y(\rho, P)} + \log f_y(\rho, P)\right], \tag{49}$$

$$\frac{\partial^2 g_y(\rho, P)}{\partial \rho^2} = \frac{\partial g_y(\rho, P)}{\partial \rho} \left[(1+\rho) \frac{\frac{\partial f_y(\rho, P)}{\partial \rho}}{f_y(\rho, P)} + \log f_y(\rho, P)\right] + g_y(\rho, P) \left[2 \frac{\frac{\partial f_y(\rho, P)}{\partial \rho}}{f_y(\rho, P)} + (1+\rho) \left\{\frac{\frac{\partial^2 f_y(\rho, P)}{\partial \rho^2}}{f_y(\rho, P)} - \left(\frac{\frac{\partial f_y(\rho, P)}{\partial \rho}}{f_y(\rho, P)}\right)^2\right\}\right], \tag{50}$$

$$\frac{\partial^3 g_y(\rho, P)}{\partial \rho^3} = \frac{\partial^2 g_y(\rho, P)}{\partial \rho^2}\left[(1+\rho)\frac{\frac{\partial f_y(\rho,P)}{\partial \rho}}{f_y(\rho,P)} + \log f_y(\rho,P)\right] + \frac{\partial g_y(\rho, P)}{\partial \rho}\left[4\frac{\frac{\partial f_y(\rho,P)}{\partial \rho}}{f_y(\rho,P)} + 2(1+\rho)\left\{\frac{\frac{\partial^2 f_y(\rho,P)}{\partial \rho^2}}{f_y(\rho,P)}\right.\right.$$
$$\left.\left.-2\left(\frac{\frac{\partial f_y(\rho,P)}{\partial \rho}}{f_y(\rho,P)}\right)^2\right\}\right] + g_y(\rho,P)\left[\left\{\frac{\frac{\partial^2 f_y(\rho,P)}{\partial \rho^2}}{f_y(\rho,P)} - \left(\frac{\frac{\partial f_y(\rho,P)}{\partial \rho}}{f_y(\rho,P)}\right)^2\right\}\left\{3 - 2\frac{\frac{\partial f_y(\rho,P)}{\partial \rho}}{f_y(\rho,P)}\right\}\right.$$
$$\left. + (1+\rho)\left\{\frac{\frac{\partial^3 f_y(\rho,P)}{\partial \rho^3}}{f_y(\rho,P)} - \frac{\frac{\partial^2 f_y(\rho,P)}{\partial \rho^2}\frac{\partial f_y(\rho,P)}{\partial \rho}}{f_y(\rho,P)^2}\right\}\right]. \tag{51}$$

For any $y \in \mathcal{Y}$, define
$$\omega_{\min}(y) := \min_{y \in \mathcal{Y}} \min_{x \in \mathcal{X}_y} W(y|x), \tag{52}$$
$$\omega_{\max}(y) := \max_{y \in \mathcal{Y}} \max_{x \in \mathcal{X}_y} W(y|x). \tag{53}$$

From (43), by using (52) and (53), we infer that
$$\frac{\partial f_y(\rho, P)}{\partial \rho} \leq \frac{f_y(\rho, P)}{(1+\rho)^2} \log \frac{1}{\omega_{\min}(y)}, \tag{54}$$
$$\frac{\partial f_y(\rho, P)}{\partial \rho} \geq \frac{f_y(\rho, P)}{(1+\rho)^2} \log \frac{1}{\omega_{\max}(y)}. \tag{55}$$

Consider any sequence $\{(\rho_k, P_k)\}_{k \geq 1}$ in $\mathbb{R}_+ \times \mathcal{P}(\mathcal{X})$ with $\mathcal{S}(P_k) \cap \mathcal{X}_y \neq \emptyset$ for all $k \in \mathbb{Z}^+$ and $(\rho_k, P_k) \to (\rho_o, P_o)$ for some $(\rho_o, P_o) \in \mathbb{R}_+ \times \mathcal{P}(\mathcal{X})$ with $\mathcal{S}(P_o) \cap \mathcal{X}_y = \emptyset$. Using (54) and (55), we deduce that

$$\mathbb{R}_+ \ni \frac{1}{(1+\rho_o)^2}\log\frac{1}{\omega_{\max}(y)} \leq \liminf_{k \to \infty} \frac{\left.\frac{\partial f_y(\rho, P_k)}{\partial \rho}\right|_{\rho=\rho_k}}{f_y(\rho_k, P_k)}$$
$$\leq \limsup_{k \to \infty} \frac{\left.\frac{\partial f_y(\rho, P_k)}{\partial \rho}\right|_{\rho=\rho_k}}{f_y(\rho_k, P_k)} \leq \frac{1}{(1+\rho_o)^2}\log\frac{1}{\omega_{\min}(y)} \in \mathbb{R}^+. \tag{56}$$

Note that (56) is evident if $\mathcal{S}(P_o) \cap \mathcal{X}_y \neq \emptyset$.

*Lemma A.1:* Given any $y \in \mathcal{Y}$, $\frac{\partial g_y(\rho, P)}{\partial \rho}$ is continuous for all $(\rho, P) \in \mathbb{R}_+ \times \mathcal{P}(\mathcal{X})$.

*Proof:* Fix any $y \in \mathcal{Y}$. Consider any $(\rho_o, P_o) \in \mathbb{R}_+ \times \mathcal{P}(\mathcal{X})$.

Note that if $\mathcal{S}(P_o) \cap \mathcal{X}_y \neq \emptyset$, then by recalling the continuity of $f_y(\cdot, \cdot)$, $\frac{\partial f_y(\rho, P)}{\partial \rho}$ and $g_y(\cdot, \cdot)$, (49) ensures that $\frac{\partial g_y(\rho, \cdot)}{\partial \rho}$ is continuous at $(\rho_o, P_o)$. Hence, suppose $\mathcal{S}(P_o) \cap \mathcal{X}_y = \emptyset$.

Let $\{(\rho_k, P_k)\}_{k \geq 1}$ be arbitrary with $\lim_{k \to \infty}(\rho_k, P_k) = (\rho_o, P_o)$. Observe that (48), along with (42) and (47), ensures that
$$\left.\frac{\partial g_y(\rho, P_k)}{\partial \rho}\right|_{\rho=\rho_k} = 0, \text{ if } \mathcal{S}(P_k) \cap \mathcal{X}_y = \emptyset. \tag{57}$$

Consider any subsequence $\{(\rho_{k_n}, P_{k_n})\}_{n \geq 1}$. Now, if all but a finite number of $P_{k_n}$ satisfy $\mathcal{S}(P_{k_n}) \cap \mathcal{X}_y = \emptyset$, then
$$\lim_{n \to \infty} \left.\frac{\partial g_y(\rho, P_{k_n})}{\partial \rho}\right|_{\rho=\rho_{k_n}} = 0, \tag{58}$$

owing to (57). Suppose this is not the case. One can verify[7] that
$$\lim_{n \to \infty} \left.\frac{\partial g_y(\rho, P_{k_n})}{\partial \rho}\right|_{\rho=\rho_{k_n}} = 0, \tag{59}$$

by using the continuity of $f_y(\cdot, \cdot)$ and $g_y(\cdot, \cdot)$, along with (46), (48), (49) and (56).

---
[7]Passing to a further subsequence $\{P_{k_{n_m}}\}_{m \geq 1}$ such that $\mathcal{S}(P_{k_{n_m}}) \cap \mathcal{X}_y \neq \emptyset$, for all $m \in \mathbb{Z}^+$, if necessary.

Combining (58) and (59), we conclude that

$$\lim_{k\to\infty} \frac{\partial g_y(\rho, P_k)}{\partial \rho}\bigg|_{\rho=\rho_k} = 0 = \frac{\partial g_y(\rho, P_o)}{\partial \rho}\bigg|_{\rho=\rho_o},$$

that implies the continuity if $\mathcal{S}(P_o) \cap \mathcal{X}_y = \emptyset$. ∎

For any $y \in \mathcal{Y}$, define

$$\overline{\omega}(y) := \max\{|\log \omega_{\min}(y)|, |\log \omega_{\max}(y)|\} \in \mathbb{R}^+, \tag{60}$$

where $\omega_{\min}(y)$ and $\omega_{\max}(y)$ are as defined in (52) and (53), respectively.

From (44), by using (60), we infer that

$$\frac{\partial^2 f_y(\rho, P)}{\partial \rho^2} \leq \frac{2 f_y(\rho, P) \log \omega_{\max}(y)}{(1+\rho)^3} + \frac{f_y(\rho, P)\overline{\omega}(y)^2}{(1+\rho)^4}, \tag{61}$$

$$\frac{\partial^2 f_y(\rho, P)}{\partial \rho^2} \geq \frac{2 f_y(\rho, P) \log \omega_{\min}(y)}{(1+\rho)^3}. \tag{62}$$

Consider any sequence $\{(\rho_k, P_k)\}_{k\geq 1}$ in $\mathbb{R}_+ \times \mathcal{P}(\mathcal{X})$ with $\mathcal{S}(P_k) \cap \mathcal{X}_y \neq \emptyset$ for all $k \in \mathbb{R}^+$ and $(\rho_k, P_k) \to (\rho_o, P_o)$ for some $(\rho_o, P_o) \in \mathbb{R}_+ \times \mathcal{P}(\mathcal{X})$ with $\mathcal{S}(P_o) \cap \mathcal{X}_y = \emptyset$. Using (61) and (62), we deduce that

$$\mathbb{R} \ni \frac{2}{(1+\rho_o)^3} \log \omega_{\min} \leq \liminf_{k\to\infty} \frac{\frac{\partial^2 f_y(\rho, P_k)}{\partial \rho^2}\big|_{\rho=\rho_k}}{f_y(\rho_k, P_k)}$$

$$\leq \limsup_{k\to\infty} \frac{\frac{\partial^2 f_y(\rho, P_k)}{\partial \rho^2}\big|_{\rho=\rho_k}}{f_y(\rho_k, P_k)} \leq \frac{2 \log \omega_{\max}(y)}{(1+\rho_o)^3} + \frac{\overline{\omega}(y)^2}{(1+\rho_o)^4} \in \mathbb{R}^+. \tag{63}$$

Note that (63) is evident if $\mathcal{S}(P_o) \cap \mathcal{X}_y \neq \emptyset$.

*Lemma A.2:* Given any $y \in \mathcal{Y}$, $\frac{\partial^2 g_y(\rho, P)}{\partial \rho^2}$ is continuous for all $(\rho, P) \in \mathbb{R}_+ \times \mathcal{P}(\mathcal{X})$.

*Proof:* Fix any $y \in \mathcal{Y}$. Consider any $(\rho_o, P_o) \in \mathbb{R}_+ \times \mathcal{P}(\mathcal{X})$.

Note that if $\mathcal{S}(P_o) \cap \mathcal{X}_y \neq \emptyset$, then, by using the continuity of $f_y(\cdot,\cdot)$, $\frac{\partial f_y(\rho,\cdot)}{\partial \rho}$, $\frac{\partial^2 f_y(\rho,\cdot)}{\partial \rho^2}$, $g_y(\cdot,\cdot)$ and $\frac{\partial g_y(\rho,\cdot)}{\partial \rho}$, (50) implies the continuity of $\frac{\partial^2 g_y(\rho,\cdot)}{\partial \rho^2}$ at the point $(\rho_o, P_o)$. Hence, suppose $\mathcal{S}(P_o) \cap \mathcal{X}_y = \emptyset$.

Let $\{(\rho_k, P_k)\}_{k\geq 1}$ be arbitrary with $\lim_{k\to\infty}(\rho_k, P_k) = (\rho_o, P_o)$. Observe that (48), along with (42) and (47), ensures that

$$\frac{\partial^2 g_y(\rho, P_k)}{\partial \rho^2}\bigg|_{\rho=\rho_k} = 0, \text{ if } \mathcal{S}(P_k) \cap \mathcal{X}_y = \emptyset. \tag{64}$$

Consider any subsequence $\{(\rho_{k_n}, P_{k_n})\}_{n\geq 1}$. Now, if all but a finite number of $P_{k_n}$ satisfy $\mathcal{S}(P_{k_n}) \cap \mathcal{X}_y = \emptyset$, then

$$\lim_{n\to\infty} \frac{\partial^2 g_y(\rho, P_{k_n})}{\partial \rho^2}\bigg|_{\rho=\rho_{k_n}} = 0, \tag{65}$$

owing to (64). Suppose this is not the case. We also have[8]

$$\lim_{n\to\infty} \frac{\partial^2 g_y(\rho, P_{k_n})}{\partial \rho^2}\bigg|_{\rho=\rho_{k_n}} = 0, \tag{66}$$

by using the continuity of $f_y(\cdot,\cdot)$, $g_y(\cdot,\cdot)$ and $\frac{\partial g_y(\rho,\cdot)}{\partial \rho}$, along with (46), (48), (49), (50), (56) and (63).

Combining (65) and (66), we conclude that

$$\lim_{k\to\infty} \frac{\partial^2 g_y(\rho, P_k)}{\partial \rho^2}\bigg|_{\rho=\rho_k} = 0 = \frac{\partial^2 g_y(\rho, P_o)}{\partial \rho^2}\bigg|_{\rho=\rho_o},$$

that implies the continuity if $\mathcal{S}(P_o) \cap \mathcal{X}_y = \emptyset$. ∎

---

[8] Passing to a further subsequence $\{P_{k_{n_m}}\}_{m\geq 1}$ such that $\mathcal{S}(P_{k_{n_m}}) \cap \mathcal{X}_y \neq \emptyset$, for all $m \in \mathbb{Z}^+$, if necessary.

Note that from (45), by using (52), (53) and (60), one can show that

$$\frac{\partial^3 f_y(\rho, P)}{\partial \rho^3} \leq \frac{1}{(1+\rho)^4} \sum_{x \in \mathcal{X}_y} P(x) W(y|x)^{\frac{1}{(1+\rho)}} \left[ 6 \log \frac{1}{\omega_{\min}(y)} - \frac{(\log \omega_{\min}(y))^3}{(1+\rho)^2} \right], \quad (67)$$

$$\frac{\partial^3 f_y(\rho, P)}{\partial \rho^3} \geq \frac{1}{(1+\rho)^4} \sum_{x \in \mathcal{X}_y} P(x) W(y|x)^{\frac{1}{(1+\rho)}} \left[ 6 \log \frac{1}{\omega_{\max}(y)} - \frac{6 \overline{\omega}(y)^2}{(1+\rho)} - \frac{(\log \omega_{\max}(y))^3}{(1+\rho)^2} \right]. \quad (68)$$

Consider any sequence $\{(\rho_k, P_k)\}_{k \geq 1}$ in $\mathbb{R}_+ \times \mathcal{P}(\mathcal{X})$ with $\mathcal{S}(P_k) \cap \mathcal{X}_y \neq \emptyset$ for all $k \in \mathbb{R}^+$ and $(\rho_k, P_k) \to (\rho_o, P_o)$ for some $(\rho_o, P_o) \in \mathbb{R}_+ \times \mathcal{P}(\mathcal{X})$ with $\mathcal{S}(P_o) \cap \mathcal{X}_y = \emptyset$. Using (67) and (68), we deduce that

$$\mathbb{R} \ni \frac{1}{(1+\rho)^4} \left[ 6 \log \frac{1}{\omega_{\max}(y)} - \frac{6 \overline{\omega}(y)^2}{(1+\rho)} - \frac{(\log \omega_{\max(y)})^3}{(1+\rho)^2} \right] \leq \liminf_{k \to \infty} \frac{\left. \frac{\partial^3 f_y(\rho, P_k)}{\partial \rho^3} \right|_{\rho=\rho_k}}{f_y(\rho_k, P_k)}$$

$$\leq \limsup_{k \to \infty} \frac{\left. \frac{\partial^3 f_y(\rho, P_k)}{\partial \rho^3} \right|_{\rho=\rho_k}}{f_y(\rho_k, P_k)} \leq \frac{1}{(1+\rho)^4} \left[ 6 \log \frac{1}{\omega_{\min}(y)} - \frac{(\log \omega_{\min(y)})^3}{(1+\rho)^2} \right] \in \mathbb{R}^+. \quad (69)$$

Note that (63) is evident if $\mathcal{S}(P_o) \cap \mathcal{X}_y \neq \emptyset$.

*Lemma A.3:* Given any $y \in \mathcal{Y}$, $\frac{\partial^3 g_y(\rho, P)}{\partial \rho^3}$ is continuous for all $(\rho, P) \in \mathbb{R}_+ \times \mathcal{P}(\mathcal{X})$.

*Proof:* Fix any $y \in \mathcal{Y}$. Consider any $(\rho_o, P_o) \in \mathbb{R}_+ \times \mathcal{P}(\mathcal{X})$.

Note that if $\mathcal{S}(P_o) \cap \mathcal{X}_y \neq \emptyset$, then, by using the continuity of $f_y(\cdot, \cdot)$, $\frac{\partial f_y(\rho, \cdot)}{\partial \rho}$, $\frac{\partial^2 f_y(\rho, \cdot)}{\partial \rho^2}$, $\frac{\partial^3 f_y(\rho, \cdot)}{\partial \rho^3}$, $g_y(\cdot, \cdot)$, $\frac{\partial g_y(\rho, \cdot)}{\partial \rho}$ and $\frac{\partial^2 g_y(\rho, \cdot)}{\partial \rho^2}$, (51) implies the continuity of $\frac{\partial^3 g_y(\rho, \cdot)}{\partial \rho^3}$ at the point $(\rho_o, P_o)$. Hence, suppose $\mathcal{S}(P_o) \cap \mathcal{X}_y = \emptyset$.

Let $\{(\rho_k, P_k)\}_{k \geq 1}$ be arbitrary with $\lim_{k \to \infty}(\rho_k, P_k) = (\rho_o, P_o)$. Observe that (48), along with (42) and (47), ensures that

$$\left. \frac{\partial^3 g_y(\rho, P_k)}{\partial \rho^3} \right|_{\rho=\rho_k} = 0, \text{ if } \mathcal{S}(P_k) \cap \mathcal{X}_y = \emptyset. \quad (70)$$

Consider any subsequence $\{(\rho_{k_n}, P_{k_n})\}_{n \geq 1}$. Now, if all but a finite number of $P_{k_n}$ satisfy $\mathcal{S}(P_{k_n}) \cap \mathcal{X}_y = \emptyset$, then

$$\lim_{n \to \infty} \left. \frac{\partial^3 g_y(\rho, P_{k_n})}{\partial \rho^3} \right|_{\rho=\rho_{k_n}} = 0, \quad (71)$$

owing to (70). Suppose this is not the case. Further, we have (passing to a further subsequence $\{P_{k_{n_m}}\}_{m \geq 1}$ such that $\mathcal{S}(P_{k_{n_m}}) \cap \mathcal{X}_y \neq \emptyset$, for all $m \in \mathbb{Z}^+$, if necessary)

$$\lim_{n \to \infty} \left. \frac{\partial^3 g_y(\rho, P_{k_n})}{\partial \rho^3} \right|_{\rho=\rho_{k_n}} = 0, \quad (72)$$

by using the continuity of $f_y(\cdot, \cdot), g_y(\cdot, \cdot), \frac{\partial g_y(\rho, \cdot)}{\partial \rho}$ and $\frac{\partial^2 g_y(\rho, \cdot)}{\partial \rho^2}$, along with (46), (48), (49), (50), (51), (56), (63) and (69).

Combining (70) and (71), we conclude that

$$\lim_{k \to \infty} \left. \frac{\partial^3 g_y(\rho, P_k)}{\partial \rho^3} \right|_{\rho=\rho_k} = 0 = \left. \frac{\partial^3 g_y(\rho, P_o)}{\partial \rho^3} \right|_{\rho=\rho_o},$$

that implies the continuity if $\mathcal{S}(P_o) \cap \mathcal{X}_y = \emptyset$.

∎

Lastly, recalling the definition of $\mathrm{E}_o(\rho, P)$ and (47), it is easy to see that

$$\mathrm{E}_o(\rho, P) = -\log \sum_{y \in \mathcal{Y}} g_y(\rho, P). \quad (73)$$

Using (73), one can check that

$$\frac{\partial \mathrm{E}_o(\rho, P)}{\partial \rho} = -\frac{\sum_{y \in \mathcal{Y}} \frac{\partial g_y(\rho, P)}{\partial \rho}}{\sum_{\tilde{y} \in \mathcal{Y}} g_{\tilde{y}}(\rho, P)}, \tag{74}$$

$$\frac{\partial^2 \mathrm{E}_o(\rho, P)}{\partial \rho^2} = -\frac{\sum_{y \in \mathcal{Y}} \frac{\partial^2 g_y(\rho, P)}{\partial \rho^2}}{\sum_{\tilde{y} \in \mathcal{Y}} g_{\tilde{y}}(\rho, P)} + \left(\frac{\partial \mathrm{E}_o(\rho, P)}{\partial \rho}\right)^2, \tag{75}$$

$$\frac{\partial^3 \mathrm{E}_o(\rho, P)}{\partial \rho^3} = -\frac{\sum_{y \in \mathcal{Y}} \frac{\partial^3 g_y(\rho, P)}{\partial \rho^3}}{\sum_{\tilde{y} \in \mathcal{Y}} g_{\tilde{y}}(\rho, P)} + 3\frac{\partial \mathrm{E}_o(\rho, P)}{\partial \rho}\frac{\partial^2 \mathrm{E}_o(\rho, P)}{\partial \rho^2} - \left(\frac{\partial \mathrm{E}_o(\rho, P)}{\partial \rho}\right)^3. \tag{76}$$

The assertions of the lemma now follow:

1) For any given $P \in \mathcal{P}(\mathcal{X})$, the concavity of $\mathrm{E}_o(\cdot, P)$ on $\mathbb{R}_+$ can either be proven by checking the non-positivity of $\frac{\partial^2 \mathrm{E}_o(\rho, P)}{\partial \rho^2}$, given in (75), or directly applying Hölder's inequality (e.g. [22, Appendix 5B]).
2) By evaluating (42), (43), (47) and (49) at $\rho = 0$ and then plugging the result into (74), one can easily check the validity of the claim.
3) By evaluating (42), (43), (44), (47), (49) and (50) at $\rho = 0$ and plugging the result into (75), one can check the validity of the claim after some algebra.
4) Fix any $P \in \mathcal{P}(\mathcal{X})$. The concavity of $\mathrm{E}_o(\cdot, P)$ on $\mathbb{R}_+$ (recall item 1) above) ensures that $\frac{\partial^2 \mathrm{E}_o(\rho, P)}{\partial \rho^2} \leq 0$, for all $\rho \in \mathbb{R}_+$. This, coupled with item 2) above, implies the claim.
5) The continuity of $g_y(\cdot, \cdot)$ on $P \in \mathcal{P}(\mathcal{X}) \times \mathbb{R}_+$ and Lemma A.1, along with (74), imply the claim.
6) The continuity of $g_y(\cdot, \cdot)$ on $P \in \mathcal{P}(\mathcal{X}) \times \mathbb{R}_+$, Lemma A.2 and item 5) above, along with (75), imply the claim.
7) The continuity of $g_y(\cdot, \cdot)$ on $P \in \mathcal{P}(\mathcal{X}) \times \mathbb{R}_+$, Lemma A.3 and items 5) and 6) above, along with (76), imply the claim.

## APPENDIX B
## PROOF OF LEMMA 3.1

The proof follows similar steps to that of [22, Theorem 5.8.5]. Let $(f, \varphi)$, $V \in \mathcal{P}(\mathcal{Y}|\mathcal{X})$ and $\delta > 0$ be as in the statement of the lemma. Define

$$G(m) := \left\{ \log \frac{V^n(\mathbf{Y}^n | \mathbf{x}^n)}{Q^n(\mathbf{Y}^n)} > n\left[\mathrm{I}(P; V) + \delta\right] \right\}, \tag{77}$$

for any $m \in \mathcal{M} := \{1, \ldots, \lceil 2^{nR} \rceil\}$, where $Q^n(\mathbf{y}^n) := \prod_{i=1}^n Q(y_i), \forall \mathbf{y}^n \in \mathcal{Y}^n$ along with $Q(y) := \sum_{x \in \mathcal{X}} P(x)V(y|x)$. Also, for the sake of notational convenience, define $i(x, y) := \log \frac{V(y|x)}{Q(y)}$, for any $(x, y) \in \mathcal{X} \times \mathcal{Y}$. Note that we have

$$\log \frac{V^n(\mathbf{y}^n | \mathbf{x}^n(m))}{Q^n(\mathbf{y}^n)} = \sum_{k=1}^n \log \frac{V(y_k | x_k(m))}{Q(y_k)} = \sum_{k=1}^n i(x_k(m); y_k), \tag{78}$$

for all $m \in \mathcal{M}$, where $\mathbf{x}^n(m)$ denotes the codeword of the code corresponding to the message $m$. Hence, for any $m \in \mathcal{M}$, we have

$$\mathrm{E}_{V^n}\left[i(\mathbf{x}^n(m), \mathbf{Y}^n) | \mathbf{x}^n(m)\right] = \sum_{k=1}^n \mathrm{E}_V\left[i(x_k(m), Y_k) | x_k(m)\right] \tag{79}$$

$$= \sum_{x \in \mathcal{X}} N(x | \mathbf{x}^n(m)) \sum_{y \in \mathcal{Y}} V(y|x) \log \frac{V(y|x)}{Q(y)} \tag{80}$$

$$= n \sum_{x \in \mathcal{X}} P(x) \sum_{y \in \mathcal{Y}} V(y|x) \log \frac{V(y|x)}{Q(y)} \tag{81}$$

$$= n\mathrm{I}(P; V), \tag{82}$$

where (79) follows from (78), (80) follows from the definition of $N(x|\mathbf{x}^n)$, which denotes the number of occurrences of the symbol $x \in \mathcal{X}$ in the string $\mathbf{x}^n$, and (81) follows from the definition of the type $P$.

Next, let $\varphi^{-1}(m) \subset \mathcal{Y}^n$ denote the decoding regions of $(f, \varphi)$, $\forall m \in \mathcal{M}$. We have

$$1 - \bar{e}(V, f, \varphi) = \sum_{m \in \mathcal{M}} \frac{1}{|\mathcal{M}|} \sum_{\mathbf{y}^n \in \varphi^{-1}(m)} V^n(\mathbf{y}^n | \mathbf{x}^n(m))$$

$$= \sum_{m \in \mathcal{M}} \frac{1}{|\mathcal{M}|} \sum_{\mathbf{y}^n \in \varphi^{-1}(m) \cap G(m)} V^n(\mathbf{y}^n | \mathbf{x}^n(m)) + \sum_{m \in \mathcal{M}} \frac{1}{|\mathcal{M}|} \sum_{\mathbf{y}^n \in \varphi^{-1}(m) \cap G(m)^c} V^n(\mathbf{y}^n | \mathbf{x}^n(m)), \quad (83)$$

Recalling (77), for any $\mathbf{y}^n \in G(m)^c$, we have

$$V^n(\mathbf{y}^n | \mathbf{x}^n(m)) \leq Q^n(\mathbf{y}^n) \exp\{n [\mathrm{I}(P; V) + \delta]\},$$

which, in turn, implies that

$$\sum_{m \in \mathcal{M}} \frac{1}{|\mathcal{M}|} \sum_{\mathbf{y}^n \in \varphi^{-1}(m) \cap G(m)^c} V^n(\mathbf{y}^n | \mathbf{x}^n(m)) \leq \sum_{m \in \mathcal{M}} \frac{1}{|\mathcal{M}|} \sum_{\mathbf{y}^n \in \varphi^{-1}(m) \cap G(m)^c} Q^n(\mathbf{y}^n) e^{n[\mathrm{I}(P; V) + \delta]}$$

$$\leq \sum_{m \in \mathcal{M}} \frac{1}{|\mathcal{M}|} \sum_{\mathbf{y}^n \in \varphi^{-1}(m)} Q^n(\mathbf{y}^n) e^{n[\mathrm{I}(P; V) + \delta]}$$

$$= \frac{\exp\{n [\mathrm{I}(P; V) + \delta]\}}{\lceil 2^{nR} \rceil} \sum_{m \in \mathcal{M}} \sum_{\mathbf{y}^n \in \varphi^{-1}(m)} Q^n(\mathbf{y}^n)$$

$$\leq \exp\{-n [R - \mathrm{I}(P; V) - \delta]\} \quad (84)$$

$$\leq e^{-n\delta}, \quad (85)$$

where (84) follows from the fact that the decoding regions are disjoint and $Q^n$ is a probability measure on $\mathcal{Y}^n$ and (85) follows from $\mathrm{I}(P; V) \leq R - 2\delta$ assumption.

Next, note that for any $m \in \mathcal{M}$

$$\sum_{\mathbf{y}^n \in \varphi^{-1}(m) \cap G_n(m)} V^n(\mathbf{y}^n | \mathbf{x}^n(m)) \leq \sum_{\mathbf{y}^n \in G_n(m)} V^n(\mathbf{y}^n | \mathbf{x}^n(m))$$

$$= V^n \{G_n(m) | \mathbf{x}^n(m)\}. \quad (86)$$

Further, using Chebyshev's inequality (recall (77), (78) and (82)), for any $m \in \mathcal{M}$ we have

$$V^n \{G_n | \mathbf{x}^n(m)\} \leq \frac{\sum_{k=1}^n \mathrm{Var}\left[i(x_k(m); Y_k) | x_k(m)\right]}{n^2 \delta^2}$$

$$= \frac{1}{n\delta^2} \left\{ \frac{1}{n} \sum_{k=1}^n \sum_{y \in \mathcal{Y}} V(y | x_k(m)) \log^2 \frac{V(y | x_k(m))}{Q(y)} - \frac{1}{n} \sum_{k=1}^n \left( \sum_{y \in \mathcal{Y}} V(y | x_k(m)) \log \frac{V(y | x_k(m))}{Q(y)} \right)^2 \right\}$$

$$\leq \frac{1}{n\delta^2} \left\{ \frac{1}{n} \sum_{k=1}^n \sum_{y \in \mathcal{Y}} V(y | x_k(m)) \log^2 \frac{V(y | x_k(m))}{Q(y)} - \left( \frac{1}{n} \sum_{k=1}^n \sum_{y \in \mathcal{Y}} V(y | x_k(m)) \log \frac{V(y | x_k(m))}{Q(y)} \right)^2 \right\} \quad (87)$$

$$= \frac{1}{n\delta^2} \left\{ \sum_{x \in \mathcal{X}} P(x) \sum_{y \in \mathcal{Y}} V(y | x) \log^2 \frac{V(y | x)}{Q(y)} - \left( \sum_{x \in \mathcal{X}} P(x) \sum_{y \in \mathcal{Y}} V(y | x) \log \frac{V(y | x)}{Q(y)} \right)^2 \right\} \quad (88)$$

$$= \frac{\mathrm{Var}\left[\log \frac{V(Y|X)}{Q(Y)}\right]}{n\delta^2}, \quad (89)$$

where (87) follows from Jensen's inequality and (88) follows from the definition of $P$. Plugging (89) into (86) and recalling (27) yields

$$\forall m \in \mathcal{M}, \quad \sum_{\mathbf{y}^n \in \varphi^{-1}(m) \cap G_n(m)} V^n(\mathbf{y}^n | \mathbf{x}^n(m)) \leq \frac{A}{n\delta^2}. \quad (90)$$

Plugging (85) and (90) into (83), we deduce that

$$\bar{e}(V, f, \varphi) \geq 1 - \frac{A}{n\delta^2} - e^{-n\delta}$$

that is (28), which was to be shown.

## APPENDIX C
## PROOF OF PROPOSITION 3.1

Let $P_n$ and $\rho_n$ achieve the maxima in (36) at rate $C - \delta_n$, i.e.,

$$E_{SP}(C - \delta_n, W) = -\rho_n(C - \delta_n) + E_o(\rho_n, P_n).$$

Now $E_{SP}(C - \delta_n, W) > 0$ for all $n$, which is evident from (4). This implies that $\rho_n > 0$ for all $n$. Since $E_o(\rho, P)$ is concave in $\rho$, it follows that

$$C - \delta_n = \left.\frac{\partial E_o(\rho, P_n)}{\partial \rho}\right|_{\rho=\rho_n} \tag{91}$$

for all $n$.

Our proof of Proposition 3.1 will use the following lemma.

*Lemma C.1:* (a) Any limit point of $\{P_n\}$ is capacity achieving.
(b) $\lim_{n \to \infty} \rho_n = 0$.
(c) $\limsup_{n \to \infty} \frac{\rho_n}{\delta_n} \leq \frac{1}{\sigma^2(W)}$.

*Proof:* Consider arbitrary subsequences $\{P_{n_k}\}_{k \geq 1}$ and $\{\rho_{n_k}\}_{k \geq 1}$ and note that, owing to the compactness of $\mathcal{P}(\mathcal{X})$ and $[0, 1]$ (switching to a further subsequence, if necessary), we may assume that

$$\lim_{k \to \infty} P_{n_k} = P_0, \quad \lim_{k \to \infty} \rho_{n_k} = \rho_0, \tag{92}$$

for some $P_0 \in \mathcal{P}(\mathcal{X})$ and $\rho_0 \in [0, 1]$.

Now (91) and part 5) of Lemma 2.1 together imply that

$$C = \left.\frac{\partial E_o(\rho, P_0)}{\partial \rho}\right|_{\rho=\rho_0}.$$

On the other hand, part 4) of Lemma 2.1 implies that

$$\left.\frac{\partial E_o(\rho, P_0)}{\partial \rho}\right|_{\rho=\rho_0} \leq I(P_0; W) \leq C. \tag{93}$$

It follows that $P_0$ is capacity achieving. Since the subsequence was arbitrary, this establishes (a).

Since $P_0$ is capacity achieving, the assumption that $\sigma^2(W) > 0$ implies that $\left.\frac{\partial^2 E_o(\rho, P_0)}{\partial \rho^2}\right|_{\rho=0} < 0$ by part 3) of Lemma 2.1. Then items 1) and 2) of Lemma 2.1 imply that the first inequality in (93) holds with equality if and only if $\rho_0 = 0$. Since the subsequence was arbitrary, this establishes (b).

Next consider $\frac{\partial E_o(\rho, P_{n_k})}{\partial \rho}$, viewed as a function of $\rho$. This function equals $I(P_{n_k}; W)$ at $\rho = 0$ by part 2) of Lemma 2.1, and it equals $C - \delta_{n_k}$ at $\rho_{n_k}$ by (91). It is differentiable in $\rho$ by part 6) of Lemma 2.1. Thus by the mean value theorem, there must exist a $\hat{\rho}_{n_k}$ in $[0, \rho_{n_k}]$ such that

$$-\left.\frac{\partial^2 E_o(\rho, P_{n_k})}{\partial \rho^2}\right|_{\rho=\hat{\rho}_n} = \frac{I(P_{n_k}; W) - C + \delta_{n_k}}{\rho_{n_k}}$$

$$\leq \frac{\delta_{n_k}}{\rho_{n_k}}.$$

Now by parts 3) and 6) of Lemma 2.1,

$$\lim_{k \to \infty} \left.\frac{\partial^2 E_o(\rho, P_{n_k})}{\partial \rho^2}\right|_{\rho=\hat{\rho}_{n_k}} = \left.\frac{\partial^2 E_o(\rho, P_0)}{\partial \rho^2}\right|_{\rho=0} = -\sigma^2(P_0, W) \leq -\sigma^2(W).$$

Combining the last two inequalities gives

$$\limsup_{n \to \infty} \frac{\rho_{n_k}}{\delta_{n_k}} \leq \frac{1}{\sigma^2(W)}. \tag{94}$$

Since the subsequence was arbitrary, this establishes (c). ∎

We are now in a position to prove Proposition 3.1. For any sufficiently large $n$, Taylor's Theorem gives (recalling items 2) and 3) of Lemma 2.1)

$$\begin{aligned} \mathrm{E}_{\mathrm{SP}}(\mathrm{C} - \delta_n, W) &:= -\rho_n[\mathrm{C} - \delta_n] + \mathrm{E}_{\mathrm{o}}(\rho_n, P_n) \\ &= \rho_n \left[\mathrm{I}(P_n; W) - \mathrm{C} + \delta_n\right] - \frac{(\rho_n)^2}{2}\sigma^2(P_n, W) + \frac{(\rho_n)^3}{6} \left.\frac{\partial^3 \mathrm{E}_{\mathrm{o}}(\rho, P_n)}{\partial \rho^3}\right|_{\rho = \bar{\rho}_n}, \end{aligned} \tag{95}$$

for some $\bar{\rho}_n \in [0, \rho_n]$. If we use the constant $M$ defined in (23), then we eventually have

$$\mathrm{E}_{\mathrm{SP}}(\mathrm{C} - \delta_n, W) \leq \rho_n \left[\mathrm{I}(P_n; W) - \mathrm{C} + \delta_n\right] - \frac{(\rho_n)^2}{2}\sigma^2(P_n, W) + \frac{(\rho_n)^3 M}{6}.$$

Since we must have $\mathrm{I}(P_n; W) \leq \mathrm{C}$, this yields

$$\begin{aligned} \mathrm{E}_{\mathrm{SP}}(\mathrm{C} - \delta_n, W) &\leq \rho_n \delta_n - \frac{(\rho_n)^2}{2}\sigma^2(P_n, W) + \frac{(\rho_n)^3 M}{6} \\ &\leq \sup_{\rho \in \mathbb{R}_+} \left\{ \rho \delta_n - \frac{\rho^2}{2}\sigma^2(P_n, W) \right\} + \frac{(\rho_n)^3 M}{6} \\ &= \frac{\delta_n^2}{2\sigma^2(P_n, W)} + \frac{(\rho_n)^3 M}{6}. \end{aligned} \tag{96}$$

Using (96) and parts (b) and (c) of Lemma C.1, we deduce that

$$\begin{aligned} \limsup_{n \to \infty} \frac{\mathrm{E}_{\mathrm{SP}}(\mathrm{C} - \delta_n, W)}{\delta_n^2} &\leq \limsup_{n \to \infty} \frac{1}{2\sigma^2(P_n, W)} \\ &\leq \frac{1}{2\sigma^2(W)}, \end{aligned} \tag{97}$$

where (97) follows from the continuity of $\sigma^2(\cdot, W)$ on $\mathcal{P}(\mathcal{X})$ (parts 3) and 6) of Lemma 2.1), Lemma C.1(a) and the definition of $\sigma^2(W)$ (cf. (5)).